\documentstyle[preprint,aps]{revtex}
\tightenlines                                                                                                                                                               

\newcommand{\be}{\begin{equation}}
\newcommand{\ee}{\end{equation}}
\newcommand{\bea}{\begin{eqnarray}}
\newcommand{\eea}{\end{eqnarray}}
\newcommand{\ba}{\begin{array}}
\newcommand{\ea}{\end{array}}

\newcommand{\Th}{\Theta}

\newcommand{\de}{\delta}

\newcommand{\pa}{\partial}

\newcommand{\no}{\nonumber}

\begin{document}

\title{The constrained modified KP hierarchy and the\\
generalized Miura transformations}

\author{Jiin-Chang Shaw$^1$, and Ming-Hsien Tu$^2$ }

\address{
$^1$ Department of Applied Mathematics, National Chiao Tung University, \\
Hsinchu, Taiwan, Republic of China,\\
and\\
$^2$ Department of Physics, National Tsing Hua University, \\
Hsinchu, Taiwan, Republic of China
}

\date{\today}

\maketitle

\begin{abstract}
In this letter, we consider the second Hamiltonian structure of the
constrained modified KP hierarchy. After mapping the Lax operator
to a pure differential operator the second structure becomes the sum
of the second and the third  Gelfand-Dickey brackets defined by this 
differential operator. We simplify this Hamiltonian structure by factorizing
the Lax operator into linear terms.

\end{abstract}
\newpage

Classical $W$-algebras has played an important role in integrable systems \cite{BS}.
It's Adler map (see, for example, \cite{D1}) from which the $W$-algebras can be 
constructed as Poisson
bracket algebras. A typical example is the $W_n$ algebra constructed from 
the second Gelfand-Dickey (GD) structure of the $n$-th Korteweg-de Vries (KdV) 
hierarchy \cite{DIZ,D}. 
Amazingly, under factorization of the KdV-Lax operator, the second Hamiltonian
structure is transformed into a much simpler one in an appropriate space of the
modified variables. 
Thus the factorization  not only provides a Miura transformation which maps the 
$n$-th KdV hierarchy to the corresponding modified hierarchies 
but also gives a free field realization of the $W_n$ algebra. 
This is what we called  the Kupershmidt-Wilson (KW)
theorem \cite{KW,D2}. In general, the above scheme is encoded in the particular form
of the Lax operator and its associated Poisson structure. Several integrable
systems have been studied based on this scheme, such as the Kadomtsev-Petviashvili (KP)
hierarchy and its reductions \cite{C,BX,BLX,D3,Y,ANP,MR}.

In this letter, we will consider a kind of reduction of the KP hierarchy called
constrained modified KP (cmKP) hierarchy \cite{OS}. Many properties of the cmKP
have been studied, such as bi-Hamiltonian structure \cite{OS},  B\"acklund transformation \cite{ST}
, modification \cite{Liu},  and conformal property \cite{HSY}, etc. 
However, a clear and conclusive
statement about the associated Poisson structure is still lacking. In the following, we will concentrate
on this problem. Especially, we will show that 
there is an interesting property of the second Poisson structure of the cmKP hierarchy
under factorization of the Lax operator into linear terms.

The cmKP hierarchy\cite{OS} has the Lax operator of the form
\be
K_n=\pa ^n+v_1\pa ^{n-1}+\cdots+v_n+\pa^{-1}v_{n+1}
\label{laxK}
\ee 
which satisfies the hierarchy equations
\be
\pa_kK_n=[(K_n^{k/n})_{\geq 1}, K_n].
\label{heK}
\ee
The second Poisson bracket associated with the Lax operator was obtained
by Oevel and Strampp \cite{OS} as follows
\be
\{F,G\}=
\int res(\frac{\de F}{\de K_n}\Th_2(\frac{\de G}{\de K_n}))
\label{poK}
\ee
where $F$ and $G$ are functionals of $K_n$  and  
\bea
\Th_2(\frac{\de G}{\de K_n})
&=&(K_n\frac{\de G}{\de K_n})_+K_n-K_n(\frac{\de G}{\de K_n}K_n)_+
+[K_n,(K_n\frac{\de G}{\de K_n})_0]
+\pa^{-1}res[K_n,\frac{\de G}{\de K_n}]K_n\no \\
&&+[K_n, \int^x(res[K_n, \frac{\de G}{\de K_n}])]
\label{hsK}
\eea
with
\be
\frac{\de G}{\de K_n}\equiv
\frac{\de G}{\de v_{n+1}}+\pa^{-1}\frac{\de G}{\de v_n}+\cdots+
\pa^{-n}\frac{\de G}{\de v_1}.
\label{varK}
\ee

Recently, Liu Q P \cite{Liu} conjectured that if the Lax operator $K_n$ is factorized as
\be
K_n=\pa^{-1}(\pa-w_1)\cdots(\pa-w_{n+1})
\label{facK}
\ee
then in terms of $\{w_i\}$ the Poisson structure (\ref{poK}) can be simplified to 
\be
\{w_i(x), w_j(y)\}=(1-\de_{ij})\de'(x-y).
\label{kwpoK}
\ee
where $\de' (x-y)\equiv \pa_x\de(x-y)$. The cases for $n=1$ and $n=2$ have been
explicitly demonstrated in \cite{Liu}. However, to the best of our knowledge, a general
proof for all $n$ is still lacking. It is the main purpose of this letter to give
an elegant and simple proof for the general case.

To simplify the Hamiltonian structure (\ref{hsK}) let us consider the operator
\bea
L_{n+1}&\equiv&\pa K_n= \pa^{n+1}+v_1\pa^{n}+(v_2+v_1')\pa^{n-1}
+\cdots+(v_{n+1}+v_n') \no \\
&\equiv&\pa^{n+1}+u_1\pa^{n}+u_2\pa^{n-1}+\cdots+u_{n+1}
\label{laxL}
\eea
which is a pure differential operator and the variables $\{v_i\}$ and $\{u_i\}$ are related by
\bea
v_1&=&u_1,\no\\
v_2&=&u_2-u_1',\no\\
&\vdots&\no\\
v_{n+1}&=&u_{n+1}-u_n'+\cdots(-1)^nu_1^{(n)}.
\eea

{\bf Proposition 1 \cite{HSY}:\/} With respect to the pure differential operator $L_{n+1}$, 
the second Poisson
bracket (\ref{poK}) now becomes
\be
\{F,G\}=
\int res(\frac{\de F}{\de L_{n+1}}\Omega(\frac{\de G}{\de L_{n+1}}))
\label{poL}
\ee
where
\be
\Omega(\frac{\de G}{\de L_{n+1}})=
(L_{n+1}\frac{\de G}{\de L_{n+1}})_+L_{n+1}-L_{n+1}(\frac{\de G}
{\de L_{n+1}}L_{n+1})_++
[L_{n+1}, \int^x(res[L_{n+1}, \frac{\de G}{\de L_{n+1}}])].
\label{hsL}
\ee
with 
\be
\frac{\de G}{\de L_{n+1}}\equiv
\pa^{-1}\frac{\de G}{\de u_{n+1}}+\pa^{-2}\frac{\de G}{\de u_n}+\cdots+
\pa^{-n-1}\frac{\de G}{\de u_1}.
\label{varL}
\ee

Besides the standard second GD structure, 
 the last piece of (\ref{hsL}) is called the third GD bracket which is compatible with
the second one \cite{DIZ}.  
Hence, under the mapping (\ref{laxL}), the Hamiltonian structure (\ref{hsK}) has 
been mapped to the sum
of the second and the third GD structure defined by the differential  operator 
$L_{n+1}$.

Now we want to show that this Hamiltonian structure can be simplified via
the following factorization
\be
L_{n+1}=(\pa-w_1)(\pa-w_2)\cdots(\pa-w_{n+1}).
\label{facL}
\ee
This yields an expression for each $u_i$ (and hence $v_i$) as a differential 
polynomial in $\{w_i\}$
(the inverse statement is not true). For example
\bea
u_1&=&-(w_1+\cdots+w_{n+1}) \no\\
u_2&=&\sum_{i<j}w_iw_j-\sum_{i=0}^{n-1}(n-i)w_{n+1-i}'          
\label{miura}                            \\
&\vdots &\no
\eea
etc. 
The expression (\ref{miura}) is called the Miura transformation. 

{\bf Proposition 2:\/} 
Under the factorization (\ref{facL}), the Poisson structure (\ref{poL}) becomes
\be
\{F, G\}=\sum_{i \neq j}\int(\frac{\de F}{\de w_i})(\frac{\de G}{\de w_j})'
\label{kwK}
\ee
 i.e., the basic building blocks $\{w_i\}$ satisfy (\ref{kwpoK}).

{\it Proof\/}: First, thanks to the KW theorem \cite{KW,D2} 
for the second GD structure, 
the first two terms of the Poisson bracket  (\ref{poL}) can be simplified as follows
\be
\{F, G\}^{GD}_2=-\sum_{i=1}^{n+1}\int(\frac{\de F}{\de w_i})(\frac{\de G}{\de w_i})', 
\label{kwL}
\ee
or
\be
\{w_i(x), w_j(y)\}^{GD}_2=-\de_{ij}\de'(x-y).
\label{kwpoL}
\ee
Thus the remaining tasks are to verify
\be
\int res(\frac{\de F}{\de L_{n+1}}[L_{n+1}, \int^x res[L_{n+1}, \frac{\de G}{\de L_{n+1}}]])
=\sum_{i,j=1}^{n+1}\int(\frac{\de F}{\de w_i})(\frac{\de G}{\de w_j})'.
\label{third}
\ee
 Let $l_i\equiv (\pa-w_i)$, then $L_{n+1}=l_1l_2\cdots l_{n+1}$ and
\bea
\int res(\frac{\de F}{\de L_{n+1}}\de L_{n+1})&=&
-\int res(\frac{\de F}{\de L_{n+1}}\sum_{i=1}^{n+1}l_1\cdots l_{i-1}\de w_il_{i+1}
\cdots l_{n+1})\no \\
&=&-\sum_{i=1}^{n+1}\int res(l_{i+1}\cdots l_{n+1} \frac{\de F}{\de L_{n+1}}l_1
\cdots l_{i-1})\de w_i \\
&=&\sum_{i=1}^{n+1}\int \frac{\de F}{\de w_i}\de w_i\no
\eea
which implies
\be
\frac{\de F}{\de w_i}=-res(l_{i+1}\cdots l_{n+1} \frac{\de F}{\de L_{n+1}}l_1\cdots l_{i-1}).
\label{id}
\ee
Now
\bea
(\sum_{i=1}^{n+1}\frac{\de F}{\de w_i})'
&=&-[\pa, res(\sum_{i=1}^{n+1}l_{i+1}\cdots l_{n+1} \frac{\de F}{\de L_{n+1}}l_1
\cdots l_{i-1})] \no\\
&=&-\sum_{i=1}^{n+1}res([\pa, l_{i+1}\cdots l_{n+1} \frac{\de F}{\de L_{n+1}}l_1
\cdots l_{i-1}]) \no\\
&=&-\sum_{i=1}^{n+1}res([l_i, l_{i+1}\cdots l_{n+1} \frac{\de F}{\de L_{n+1}}l_1
\cdots l_{i-1}]) \no\\
&=&-res[L_{n+1}, \frac{\de F}{\de L_{n+1}}].
\eea
Hence,
\be
\sum_{i=1}^{n+1}\frac{\de F}{\de w_i}
=-\int^x res[L_{n+1}, \frac{\de F}{\de L_{n+1}}].
\label{id3}
\ee
Note that we have substituted $l_i$ for $\pa$ in the third line because nothing will change.

Therefore, 
\bea
\sum_{i,j=1}^{n+1}\int(\frac{\de F}{\de w_i})(\frac{\de G}{\de w_j})'
&=&-\int (\sum_{i=1}^{n+1}\frac{\de F}{\de w_i})'(\sum_{j=1}^{n+1}\frac{\de G}
{\de w_j})\no\\
&=&-\int res([L_{n+1}, \frac{\de F}{\de L_{n+1}}]\int^x res[L_{n+1}, \frac{\de G}
{\de L_{n+1}}])\no\\
&=&\int res(\frac{\de F}{\de L_{n+1}}
[L_{n+1}, \int^x res[L_{n+1}, \frac{\de G}{\de L_{n+1}}]]).\mbox{\qquad $\Box$}
\label{id2}
\eea

{\bf Proposition 3:\/} If the Hamiltonian $H_k$ of the cmKP hierarchy equations
$\pa_kK_n=[(K_n^{k/n})_{\geq 1}, K_n]=\Th_2(\frac{\de H_k}{\de K_n})$
with respect to the second structure is expressed in terms of $\{w_i\}$
by the Miura transformation, then the corresponding modified equations
will be
\be
\pa_k w_i=\sum_{j\neq i}(\frac{\de H_k}{\de w_j})'.
\ee
{\it Proof\/}: This is just a corollary of the Proposition 2. \qquad $\Box$

Finally, we would like to provide another interesting property of the Poisson 
structure (\ref{poL})  although it is less relevant to the present case. 
In fact, it has been shown \cite{T} that the Poisson structure  (\ref{poL})
can be associated to the  operator of the form
\be
L=\pa^N+u_1\pa^{N-1}+\cdots+u_N+\sum_{i=1}^M\phi_i\pa^{-1}\psi_i.
\ee
Therefore we can discuss the Poisson structure (\ref{poL}) under the
factorization of the Lax operator containing inverse linear terms.

{\bf Proposition 4:\/}  Let $L$ be a pseudo-differential operator
of order $n-m$. If $L$ admits the following factorization
 (generalized Miura transformation)
\be
L=(\pa-a_1)\cdots(\pa-a_n)(\pa-b_1)^{-1}\cdots(\pa-b_m)^{-1}
\label{gfacL}
\ee
then the Poisson structure (\ref{poL}) associated with $L$ becomes
\bea
\{a_i(x), a_j(y)\}&=&(1-\de_{ij})\de'(x-y), \no\\
\{b_i(x), b_j(y)\}&=&(1+\de_{ij})\de'(x-y),
\label{gkwb}\\
\{a_i(x), b_j(y)\}&=&\de'(x-y).\no
\eea
{\it Proof\/}: It has been shown \cite{D3,Y,ANP,MR} that the second GD 
bracket with respects
to the factorization (\ref{gfacL}) are given by
\bea
\{a_i(x),a_j(y)\}^{GD}_2&=&-\de_{ij}\de'(x-y),\no\\
\{b_i(x),b_j(y)\}^{GD}_2&=&\de_{ij}\de'(x-y),\\
\{a_i(x),b_j(y)\}^{GD}_2&=&0.\no
\eea
Hence, we only need to treat the third structure and to show that
\be
\int res(\frac{\de F}{\de L}[L, \int^x res[L, \frac{\de G}{\de L}]])
=\int(\sum_{i=1}^n\frac{\de F}{\de a_i}+\sum_{j=1}^m\frac{\de F}{\de b_j})
(\sum_{i=1}^n\frac{\de G}{\de a_i}+\sum_{j=1}^m\frac{\de G}{\de b_j})'.
\ee 
Let $A_i=(\pa-a_i)$ and  $B_j=(\pa-b_j)$ then
\bea
\de F
&=&\int res(\frac{\de F}{\de L}\de L)\no\\
&=&\int res(\frac{\de F}{\de L}\sum_{i=1}^nA_1\cdots A_{i-1}\de A_i
\cdots A_nB_1^{-1}\cdots B_m^{-1})\no\\
& &+\int res(\frac{\de F}{\de L}A_1\cdots A_n\sum_{j=1}^mB_1^{-1}
\cdots B_{j-1}^{-1}
\de B_j^{-1}\cdots B_m^{-1})
\label{rela1}\\
&\equiv&\int(\sum_{i=1}^n\frac{\de F}{\de a_i}\de a_i+\sum_{j=1}^m
\frac{\de F}{\de b_j}\de b_j).
\label{rela2}
\eea
Substituting $\de A_i=-\de a_i$ and $\de B_j^{-1}=B_j^{-1}\de b_jB_j^{-1}$ 
into (\ref{rela1})
and comparing with (\ref{rela2}), we obtain
\bea
\frac{\de F}{\de a_i}&=&-res(A_{i+1}\cdots A_nB_1^{-1}\cdots B_m^{-1}
\frac{\de F}{\de L}A_1\cdots A_{i-1})\\
\frac{\de F}{\de b_j}&=&res(B_j^{-1}\cdots B_m^{-1}\frac{\de F}{\de L}A_{1}
\cdots A_nB_1^{-1}\cdots B_j^{-1}).
\eea
Thus
\bea
\sum_{i=1}^n(\frac{\de F}{\de a_i})'+\sum_{j=1}^m(\frac{\de F}{\de b_j})'
&=&\sum_{i=1}^n[\pa, \frac{\de F}{\de a_i}]+\sum_{j=1}^m[\pa, \frac{\de F}
{\de b_j}]\no\\
&=&-\sum_{i=1}^n res[A_i, A_{i+1}\cdots A_nB_1^{-1}\cdots B_m^{-1}
\frac{\de F}{\de L}A_1\cdots A_{i-1}]\no\\
& &+\sum_{j=1}^m res[B_j, B_j^{-1}\cdots B_m^{-1}\frac{\de F}{\de L}A_1
\cdots A_nB_1^{-1}\cdots B_j^{-1}]\no\\
&=&-res[L, \frac{\de F}{\de L}]
\eea
which implies
\be
\sum_{i=1}^n\frac{\de F}{\de a_i}+\sum_{j=1}^m\frac{\de F}{\de b_j}
=-\int^x res[L, \frac{\de F}{\de L}].
\ee
Now 
\bea
\int(\sum_{i=1}^n\frac{\de F}{\de a_i}+\sum_{j=1}^m\frac{\de F}{\de b_j})
(\sum_{i=1}^n\frac{\de G}{\de a_i}+\sum_{j=1}^m\frac{\de G}{\de b_j})'
&=&-\int(\sum_{i=1}^n\frac{\de F}{\de a_i}+\sum_{j=1}^m\frac{\de F}{\de b_j})'
(\sum_{i=1}^n\frac{\de G}{\de a_i}+\sum_{j=1}^m\frac{\de G}{\de b_j})\no\\
&=&-\int res([L, \frac{\de F}{\de L}]\int^x res[L, \frac{\de G}{\de L}])\no\\
&=&\int res(\frac{\de F}{\de L}[L, \int^x res[L, \frac{\de G}{\de L}]]). 
\mbox{\qquad $\Box$}
\eea
In summary, we have shown that the second Hamiltonian structure of the 
cmKP hierarchy has a very simple realization. In terms of the variables
$\{w_i\}$, the Lax operator $K_n$ can be factorized as (\ref{facK})
and the Poisson structure (\ref{poK}) is mapped into a much
simpler form (\ref{kwpoK}). We also discuss the Poisson structure
(\ref{poL}) under factorization of the Lax operator  containing inverse
linear terms.  The resulting brackets (\ref{gkwb}) turns out to be simple as well.
We hope that we can explore the usage of these brackets in the future.

{\it Note added\/}: after submitting this manuscript we note that similar results are also
obtained by Q. P. Liu \cite{L2}

{\bf Acknowledgments\/}
We would like to thank Professor W-J Huang for inspiring discussions
and Dr. M-C Chang for reading the manuscript. 
This work was supported by the National Science Council of the
Republic of China under grant No. NSC-86-2112-M-007-020.


\end{document}